\documentstyle[epsbox]{jpsj}

\def\vereq#1#2{\lower 3pt\vbox {\baselineskip 1.5pt \lineskip 1.5pt \ialign
{$\m@th #1\hfill ##\hfil $\crcr #2\crcr \sim \crcr }}}
\def\m@th{\mathsurround=0pt}
\def\mi{\mbox{i}}
\def\md{\mbox{d}}
\def\mibp{\mib{p}}
\def\mibpp{\mib{p'}}

\def\mibq{\mib{q}}

\def\runtitle{
Vertex Corrections in 
the Spin-fluctuation-induced Superconductivity
}
\def\runauthor{Takuya OKABE}

\title{
Vertex Corrections in 
the Spin-fluctuation-induced Superconductivity
}

\author
{ 
Takuya {\sc Okabe}\footnote{
E-mail: okabe@phys.eg.gunma-u.ac.jp}
}

\inst
{
Faculty of Engineering, Gunma University, Kiryu, Gunma 376-8515.
}

\recdate
{June 28, 1999}

\abst
{
We evaluate vertex corrections to $T_c$
on the basis of the antiferromagnetic spin-fluctuation model of
the high-$T_c$ superconductivity.
It is found that 
the corrections are attractive in the $d_{x^2-y^2}$ channel,
and they become appreciable 
as we go through an intermediate-coupling regime of $T_c\simeq 100$K,
the maximum $T_c$ attainable
in the one-loop \'Eliashberg calculation.
}

\kword
{high-$T_c$ superconductor, antiferromagnetic spin
fluctuation, vertex correction
}

\begin{document}
\sloppy
\twocolumn
\maketitle

As a model for the high-$T_c$ superconductivity,
the spin fluctuation mechanism
has been one of the most widely discussed.~\cite{MTU, MBP}\
The model assumes that quasiparticle is coupled with
antiferromagnetic spin fluctuation, represented by
a peculiar low-energy expression for the magnetic
susceptibility.~\cite{MTU, MMP}\
The phenomenological coupling constant
to fit the transition temperature $T_c$
is
used to explain, among others,
the anomalous transport properties consistently.~\cite{MTU,
MP1,radtke,MP2,MP3,HR,SN,SP,YY,KK,DK}\
On the other side, from a microscopic point of view,
numerical studies based on the fluctuation exchange (FLEX)
approximation~\cite{flex}\
have been carried out by many authors
to estimate $T_c$ as well as
to explain the deviations from the normal Fermi liquid
behavior.~\cite{flex,hub,dp,KKU}\
Computational feasibility of these strong coupling theories
rests on the effective use of a fast Fourier transform (FFT) 
algorithm.
To asses the quantitative aspect of the theories,
corrections coming from higher-order terms
are investigated for
the vertex function at some fixed external momenta,
e.g., on the basis of the spin-fluctuation model,
and the qualitative features of the effect are emerging to
some extent.~\cite{GC,JRS,AVC,AIM,AS,CMM,PM}\
However, the total effect of the vertex corrections
on the physical observables
is yet to be estimated numerically.
Indeed to do this is generally formidable
because of the inapplicability of the
FFT to a required additional sum on internal frequency and momentum.
In this paper,
we manage to evaluate the vertex
corrections to $T_c$, and discuss questions of convergence 
of the formal perturbation theory with respect to the
coupling constant.

To put it concretely,
the following investigation is based on the model of
Monthoux and Pines (MP),~\cite{MP1,MP2,MP3}
in which
the self-energy $\Sigma(\mibp,\mi\omega_n)$
is determined as a self-consistent solution of the equations, 
\begin{equation}
  \Sigma(\mibp,\mi\omega_n)=
  g^2\frac{T}{\Omega}
  \sum_{q, m}
  \chi(\mibq,\mi\nu_m) G(\mibp-\mibq,\mi\omega_n-\mi\nu_m),
  \label{Sigma=g2chiG}
\end{equation}
and
\begin{equation}
  G(\mibp,\mi\omega_n)^{-1}
  =  G^{(0)}(\mibp,\mi\omega_n) ^{-1}
  -  \Sigma(\mibp,\mi\omega_n) +\delta \mu,
  \label{G-1=G0-1-Sigma}
\end{equation}
where 
\begin{equation}
  G^{(0)}(\mibp,\mi\omega_n)=\frac{1}{\mi\omega_n-\varepsilon_p+\mu^{(0)}},
\end{equation}
and
\begin{equation}
  \varepsilon_p=-2t(\cos p_x+\cos p_y)-4t'\cos p_x \cos p_y.
  \label{varepsilon}
\end{equation}
In eq.~(\ref{Sigma=g2chiG}), $\chi(\mibq,\mi\nu_m)$
as a function of the Matsubara frequency $\nu_m=2m\pi T$
is inferred from
the low-energy form of the magnetic susceptibility,~\cite{MMP}
\begin{equation}
 \chi(\mibq,\omega )=
\frac{\chi_Q \omega_{\rm sf}}{\omega_q-\mi\omega},
\label{chiqomega}
\end{equation}
where
\begin{equation}
\omega_q\equiv\omega_{\rm sf}  (1+\xi^2(\mibq-\mib{Q})^2),
\qquad \mib{Q}=(\pi,\pi),
\end{equation}
for $q_x>0$ and $q_y>0$.
We assume
\begin{eqnarray}
  \chi(\mibq,\mi\nu_n)
   &=&
  -\frac{1}{\pi}
  \int^{\omega_0}_{-\omega_0}
  \frac{\mbox{Im}\chi(\mibq,\omega )}{\mi\nu_n-\omega}
  {\md \omega}
  \nonumber\\  &=&\displaystyle
  \frac{2}{\pi}
  \int^{\omega_0}_0
  \frac{\omega
  \mbox{Im}\chi(\mibq,\omega )}{\nu_n^2+\omega^2}
{\md \omega} \qquad (n\ne 0)  \nonumber\\
&=& \chi(\mibq,\omega=0). \qquad\qquad\qquad\,\, (n=0) 
\label{chinne0}
\end{eqnarray}
Here it is noted that a cutoff $\omega_0$ has to be
artificially introduced
so as to meet the condition
$\chi(\mibq,\mi\nu_n)\rightarrow 1/\nu_n^2$ as
$|\nu_n|\rightarrow \infty$.~\cite{MP1}\
For eq.~(\ref{chiqomega}),
the integral in eq.~(\ref{chinne0}) is analytically evaluated;
\begin{equation}
  \chi(\mibq,\mi\nu_n)=
\frac{2\chi_Q\omega_{\rm sf}/\pi}{\nu_n^2-\omega_q^2}
\left(
  |\nu_n|\tan^{-1}\frac{\omega_0}{|\nu_n|}
- \omega_q \tan^{-1}\frac{\omega_0}{\omega_q}
\right).
\end{equation}

To estimate the critical temperature $T_c$,
the linearized gap equation is used,
\begin{eqnarray}
  \mit\Phi(\mibp,\mi\omega_n)&=&
  -\frac{T}{\Omega}
  \sum_{p', n'}
  V
  (\mibp,\mi\omega_n;\mibpp,\mi\omega_{n'}) \nonumber\\
&&\times  |G(\mibpp,\mi\omega_{n'})|^2
  \mit\Phi(\mibpp,\mi\omega_{n'}),
  \label{Phi=VG2Phi}
\end{eqnarray}
\halftext
where $\Phi(\mibp,\mi\omega_n)$ is the anomalous self-energy.
The pairing potential 
$V
(\mibp,\mi\omega_n;\mibpp,\mi\omega_{n'})$
reads
\begin{eqnarray}
V
(\mibp,\mi\omega_n;\mibpp,\mi\omega_{n'})
  =V^{(1)}(\mibp-\mibpp,\mi\omega_n-\mi\omega_{n'}) \qquad  \nonumber\\
+ V_v^{(2)}(\mibp,\mi\omega_n;\mibpp,\mi\omega_{n'})
+ V_c^{(2)}(\mibp,\mi\omega_n;\mibpp,\mi\omega_{n'}),\quad
\label{Veff}
\end{eqnarray}
\halftext
where
\begin{equation}
V^{(1)}(\mibp-\mibpp,\mi\omega_n-\mi\omega_{n'})  
=  g^2 \chi(\mibp-\mibpp,\mi\omega_n-\mi\omega_{n'}).
\end{equation}
The second and third terms in eq.~(\ref{Veff})
originate from the vertex corrections that we discuss below.

As we are concerned about the $d$-wave
instability, introducing a notation
\begin{equation}
  \left\langle f(\mibp) \right\rangle_p\equiv
  \frac{1}{\Omega}\sum_{p}(\cos(p_x)-\cos(p_y))f(\mibp),
\end{equation}
we put eq.~(\ref{Phi=VG2Phi}) into
\begin{equation}
  \bar{\mit\Phi}(\mi\omega_n)=
  \sum_{n'}K(\mi\omega_n,\mi\omega_{n'})
  \bar{\mit\Phi}(\mi\omega_{n'}),
  \label{Psi=KPsi}
\end{equation}
where
\begin{eqnarray}
  \bar{\mit\Phi}(\mi\omega_n)&=&
  \left\langle\mit\Phi(\mibp,\mi\omega_n)\right\rangle_p,\\
  K(\mi\omega_n,\mi\omega_{n'})
  &=&  K^{(1)}+  K^{(2)}_v+  K^{(2)}_c,
\end{eqnarray}
and
\begin{equation}
  K^{(1,2)}_i
  =
  -T\left\langle
  V^{(1,2)}_i
  |G(\mibpp,\mi\omega_{n'})|^2
  \right\rangle_{p,p'}.
  \label{K(1)}
\end{equation}
\halftext
Here $K^{(2)}_i(\mi\omega_n,\mi\omega_{n'})$ ($i=v,c$)
come from the vertex corrections.
As is clear from eq.~(\ref{Psi=KPsi}),
the condition that the largest eigenvalue of
$K(\mi\omega_n,\mi\omega_{n'})$ reaches unity
provides a nonzero solution $\bar{\mit\Phi}(\mi\omega_n)$,
thus defines $T_c$.
Below we look for a real solution $\bar{\mit\Phi}(\mi\omega_n)$,
for which the imaginary part of
$K(\mi\omega_n,\mi\omega_{n'})$,
namely, $\mbox{Im}\, V
(\mibp,\mi\omega_n;\mibpp,\mi\omega_{n'})$,
is neglected.

As for the parameters,
we assume $t=0.25$eV and $t'=-0.45t$
for eq.~(\ref{varepsilon}),
and 
we take
\begin{equation}
\xi=2.3,\, \chi_Q=75 \mbox{/eV},\, \omega_{\rm sf}=14\mbox{meV},
\end{equation}
to describe $\chi(\mibq, \omega)$
of YBa$_2$Cu$_3$O$_7$ at $T_c=90$K,
according to MP.~\cite{MP3}\
The chemical potential $\mu^{(0)}$ is fixed by
\begin{equation}
  n=
  \frac{2T}{\Omega}\sum_{p,n}
  G^{(0)}(\mibp,\mi\omega_n)\mbox{e}^{+\mi\omega_n 0}
=
\frac{2}{\Omega}\sum_{p} \frac{1}{
\mbox{e}^{(\varepsilon_p-\mu^{(0)})/T}+1},
\end{equation}
in which we assume $n=0.75$ throughout this paper.
The shift $\delta\mu$ in eq.~(\ref{G-1=G0-1-Sigma})
is adjusted in every iteration to assure
\begin{equation}
\delta n= \frac{2T}{\Omega}\sum_{p,n}
\left(  G(\mibp,\mi\omega_n)-
  G^{(0)}(\mibp,\mi\omega_n)\right)=0,
\label{deln}
\end{equation}
as this is easier to handle than the formally divergent sum,
\(
n=
(2T/\Omega)
\sum_{p,n}G(\mibp,\mi\omega_n).
\)
For practical purposes,
the Matsubara sums in
eqs.~(\ref{Sigma=g2chiG}), (\ref{Phi=VG2Phi}) and
(\ref{deln}) are restricted within
a finite range $|\omega_n|, |\nu_m| \le \omega_{\rm c}$.
To avoid spurious temperature dependences,
the cutoff is fixed at $\omega_{\rm c}=6.2$eV
$\sim 3$ times the bandwidth,
for which we have
$|\nu_m|\simeq \omega_{\rm c}$ for $m=\pm 2^7$ at $T=90$K.~\cite{MP3}\

\begin{figure}[t]
\halftext
\centerline{\epsfile{file=fig1,width=7.cm}}
\caption{The coupling constant $g^2$ to account for $T_c=90$K
as a function of the cutoff $\omega_0$,
calculated on a 16$\times $16 lattice with periodic boundary conditions.
}
\label{fig:1}
\end{figure}
\begin{figure}[t]
\halftext
\centerline{\epsfile{file=fig2,width=7.cm}}
\caption{
Critical temperature $T_c$ is shown as a function of
$g^2$ for the cutoff $\omega_0=0.4$eV.
The other parameters are given in the text.
The results are shown for a 16$\times$16 (circles),
32$\times$32 (squares), 64$\times$64 (diamonds) and
128$\times$128 (triangles) square lattice.
The effect of $\Sigma(\mibp,\mi\omega_n)$
is not taken into account for the open symbols,
while it is included for the closed symbols.
A closed triangle for 128$\times$128 is not shown.
}
\label{fig:2}
\end{figure}
The above, except for the cutoff $\omega_0$ in eq.~(\ref{chinne0}),
are all the necessary ingredients to reproduce
the results of MP.~\cite{MP3}\
Nevertheless, we could not derive them precisely,
though qualitative features are consistently reproduced.
Let us discuss the point briefly.
First we have to note the $\omega_0$-dependence of
$T_c$ as a function of the coupling constant $g^2$.
In Fig.~\ref{fig:1},
$g^2$ to give $T_c=90$K is shown
as a function of $\omega_0$.
In this result, $K^{(1)}$ in eq.~(\ref{K(1)}) is used for
the eigenequation (\ref{Psi=KPsi}), i.e., 
the vertex correction $K^{(2)}_i$ is not taken into account,
but the effect of $\Sigma(\mibp, \mi\omega_n)$ is.
As we see from the figure,
we cannot fix $g^2$ just from $T_c(g^2)=90$K without
the knowledge of the cutoff $\omega_0$ in eq.~(\ref{chinne0}).
The strong dependence on $\omega_0$ in the low-energy
region $\omega_0\le 0.3$eV reflects 
that the transition temperature $T_c$,
unlike the transport properties
controlled by the quasiparticle damping,
is not determined solely from 
the low-energy expression, eq.~(\ref{chiqomega}).
Indeed, this is one of the general problems to
infer the full structure of $\chi(\mibq,\mi\nu_n)$
entirely from the low-energy `observable' $\chi(\mibq,\omega)$,
and to settle $\omega_0$ may be a key point in
the discrepancy between Radtke $et$ $al.$~\cite{radtke}
and MP.~\cite{SN}\
As it is not our purpose to discuss this point further,
for the time being,
we assume eq.~(\ref{chiqomega}) up to the cutoff energy $\omega_0$
in eq.~(\ref{chinne0}),
and arbitrarily
set $\omega_0=0.4$eV$\simeq 4.6\times 10^3$K, following MP.~\cite{MP3}\
This value is used throughout in the following.
Then we obtain $g^2=0.57$eV$^2$ for $T_c=90$K,
still in disagreement with $g^2=0.41$eV$^2$ of MP.~\cite{MP3}\
This is not due to the size of a square lattice,
as we see in Fig.~\ref{fig:2}, where
$T_c$ is shown as a function of $g^2$.
At $T=T_c=90$K, the 16$\times$16 lattice is large enough for 
us to conclude $g^2=0.57$eV$^2$ in the self-consistent calculation 
including the effect of $\Sigma(\mibp,\mi\omega_n)$.
A close inspection indicates that
the disagreement 
originates in details of $\chi(\mibq,\mi\nu_m)$.
In fact, we find $g^2=0.34$, smaller than $g^2=0.41$ of MP,
if we adopt the second line, instead of the third line,
of eq.~(\ref{chinne0}) for $n=0$ too.
This means that
$T_c$ depends sensitively on how we prepare
$\chi(\mibq,\mi\nu_n)$ in the low-energy regime.
This is complementary to the above remark on the high-energy
contribution to $T_c$.
As this quantitative difference is not of our primal
concern either,  deferring this problem,
we choose to use our own definition,
i.e.,  $\chi(\mibq,\mi\omega_n)$ for $n=0$ is specified
separately by eq.~(\ref{chinne0}).
The qualitative results presented below
are not affected by this choice.

\begin{figure}[t]
\halftext
\centerline{
\epsfile{file=fig3,width=7.cm}
}
\caption{The diagram (a)
for the vertex corrected pairing potential $V_v^{(2)}$
and (b) for $V_c^{(2)}$.
}
\label{fig:3}
\end{figure}
Now let us discuss how we evaluate the vertex correction.
The pairing potentials
$V^{(2)}_v(\mibp,\mi\omega_n;\mibpp,\mi\omega_{n'})$
and
$V^{(2)}_c(\mibp,\mi\omega_n;\mibpp,\mi\omega_{n'})$
including the vertex correction are
diagrammatically represented by
Fig.~\ref{fig:3}(a) and Fig.~\ref{fig:3}(b), respectively.
These potentials at low frequencies
$\omega_n=\omega_{n'}=\pi T$ (for $n=n'=0$)
are particularly studied by Monthoux.~\cite{PM}\
To see the effect on $T_c$ precisely, however,
we have to evaluate the kernel $K^{(2)}_i(\mi\omega_n,\mi\omega_{n'})$,
eq.~(\ref{K(1)}),
for a full set of the Matsubara frequencies $\omega_n$ and
$\omega_{n'}$,
then the kernel must be diagonalized.
In effect, this is not practical at present.
Thus, as a tractable method,
we set up perturbation theory
to evaluate the vertex corrections
to the eigenvalue $\kappa$ of the kernel.

We shall make effective use of the results
obtainable by means of the FFT.
Let us introduce
the eigenfunction $\bar{\mit\Phi}^{(1)}(\mi\omega_n)$
for the largest eigenvalue $\kappa^{(1)}$ 
of the kernel $K^{(1)}(\mi\omega_n,\mi\omega_{n'})$;
\(
 \bar{\mit\Phi}^{(1)}(\mi\omega_n)=
  \sum_{n'}K^{(1)}(\mi\omega_n,\mi\omega_{n'})
 \bar{\mit\Phi}^{(1)}(\mi\omega_{n'})
= \kappa^{(1)} \bar{\mit\Phi}^{(1)}(\mi\omega_n).
\)
Then the eigenvalue $\kappa$ including
the vertex corrections is given by 
\begin{equation}
  \kappa=\kappa^{(1)}+\kappa^{(2)}_v+\kappa^{(2)}_c,
\end{equation}
where
\begin{equation}
\kappa^{(2)}_i= \sum_{n,n'}
 \bar{\mit\Phi}^{(1)}(\mi\omega_{n})
  K^{(2)}_i(\mi\omega_n,\mi\omega_{n'})
 \bar{\mit\Phi}^{(1)}(\mi\omega_{n'}).
 \label{kappa2}
\end{equation}
We assume 
$\bar{\mit\Phi}^{(1)}(\mi\omega_n)$ is normalized.
On physical grounds,
the norm of $\bar{\mit\Phi}^{(1)}(\mi\omega_{n})$
decreases quite rapidly as $|\omega_{n}|$ increases.
Therefore, 
the sum over the Matsubara frequencies in eq.~(\ref{kappa2})
is allowed to be restricted in a narrow region around
$(n, n') \sim (0,0)$.
In effect, we evaluate $K^{(2)}_i(\mi\omega_n,\mi\omega_{n'})$
for a $16\times 16$ mesh around the Fermi energy.
Moreover, in the remainder of the paper,
the results are calculated on a $16\times 16$ square lattice.
Measured in terms of the weight
$\left|\bar{\mit\Phi}^{(1)}(\mi\omega_{n})\right|^2$,
we find
$
\sum_{|\omega_n|\le 15\pi T}
\left|\bar{\mit\Phi}^{(1)}(\mi\omega_{n})\right|^2=$
0.98, 0.91  and 0.78
at $T=T_c=$90K, 45K and 22K, respectively.
Even at low $T_c$,
the error involved is not appreciable,
for the coupling constant itself is small there.
As Fig.~\ref{fig:2} shows,
a $16\times16$ mesh in the momentum space
is large enough to
grasp the qualitative features caused by the vertex corrections.

\begin{figure}[t]
\halftext
\centerline{\epsfile{file=fig4,width=7cm}}
\caption{$T_c$ as a function of $g^2$.
Triangles; calculated without $\Sigma(\mibp, \mi\omega_n)$.
Circles; including the effect of $\Sigma(\mibp,\mi\omega_n)$.
Diamonds; including the effect of $\Sigma(\mibp,
\mi\omega_n)$ as well as the vertex corrections $V_v^{(2)}$
and $V_c^{(2)}$.
Two squares at $T=90$K and 45K
are calculated with 
$\Sigma(\mibp,\mi\omega_n)$ and $V_c^{(2)}$
but without $V_v^{(2)}$. 
}
\label{fig:4}
\end{figure}
To prepare
$V^{(2)}_i(\mibp,\mi\omega_n;\mibpp,\mi\omega_{n'}) $
is
most time-consuming.
Therefore, first 
we use the bare Green's function $G^{(0)}(\mibp,\mi\omega_{n})$
instead of $G(\mibp,\mi\omega_{n})$ 
to provide $V^{(2)}_i(\mibp,\mi\omega_n;\mibpp,\mi\omega_{n'})$.
With  $\left  \langle
  V^{(2)}_i(\mibp,\mi\omega_n;\mibpp,\mi\omega_{n'})  \right
\rangle_{p}$ thus calculated beforehand and 
$G(\mibp,\mi\omega_{n})$ of the solution of
eqs.~(\ref{Sigma=g2chiG}) and (\ref{G-1=G0-1-Sigma}),
we calculate $K^{(2)}_i(\mi\omega_n,\mi\omega_{n'})$
in eq.~(\ref{K(1)}).
Then,
to evaluate $\kappa^{(2)}_i$, eq.~(\ref{kappa2}), is
straightforward,
and the critical coupling $g^2$ at $T_c$ is determined.
Results thus obtained are shown in Fig.~\ref{fig:4},
where the triangles (without $\Sigma(\mibp,\mi\omega_n)$)
and circles (with $\Sigma(\mibp,\mi\omega_n)$)
denote the results without the vertex corrections. (See Fig.~\ref{fig:2}).
The diamonds include the vertex correction
$V_v^{(2)}$ as well as $V_c^{(2)}$,
while
only the effect of $V_c^{(2)}$ is taken into account
for the two squares at $T=90$K and 45K.

Several points are noted from the figure.
In the first place,
both the effects of $V_v^{(2)}$ and $V_c^{(2)}$ 
are attractive on the whole,
or enhances $T_c$ of the $d$-wave instability.
The effect of $V_c^{(2)}$ (Fig.~\ref{fig:3}(b)),~\cite{JRS}\
however, is negligibly small,
as noted by Monthoux.~\cite{PM}\
On the other hand,
the effect of $V_v^{(2)}$ (Fig.~\ref{fig:3}(a))
is prominent.
In particular, it affects the result of MP,
denoted by the circles interpolated with the solid line in
Fig.~\ref{fig:4},
that the maximum transition temperature
attainable in this model is about 100K.~\cite{MP1}\
In fact, $T_c$  as a function of $g^2$
shows no sign of saturation,
and keeps increasing beyond 200K
when the vertex correction $V_v^{(2)}$ is taken into account.
In this regard, 
the vertex correction has an effect more than
a mere scale-up of the effective coupling constant $g^2$.

\begin{figure}[t]
\halftext
\centerline{\epsfile{file=fig5,width=7.cm}}
\caption{
At $T=90$K, $\kappa^{(1)}$ and
$\kappa^{(1)}+\kappa^{(2)}$ are shown as a function of
$g^2$.
Squares, including only the effect of $V_c^{(2)}$,
are overlapping with circles to denote $\kappa^{(1)}$
without the vertex corrections.
}
\label{fig:5}
\end{figure}
Next, the effect of $\Sigma(\mibp,\mi\omega_n)$
on $V_i^{(2)}$ has to be investigated.
To this end, $G(\mibp,\mi\omega_{n})$ to meet
eqs.~(\ref{Sigma=g2chiG}) and (\ref{G-1=G0-1-Sigma})
is used to evaluate
$V^{(2)}_i(\mibp,\mi\omega_n;\mibpp,\mi\omega_{n'})$.
The maximum eigenvalues calculated for $T=90$K are
shown in Fig.~\ref{fig:5} as a function of $g^2$.
The effect of $\Sigma(\mibp,\mi\omega_n)$ is
to weaken the vertex corrections.
The effect, however, is not appreciable for $T_c=90$K,
as we see from Fig.~\ref{fig:5} in which we see $g^2=0.36$ while
we have $g^2=0.32$ in Fig.~\ref{fig:4} in the case
including the vertex corrections.
Comparing these with $g^2=0.57$
without the vertex corrections,
we conclude that 
the correction due to $\Sigma(\mibp,\mi\omega_n)$ in
$V_i^{(2)}$ is not important at least at $T_c=90$K.
In other words,
if the coupling constant $g$ should be evaluated to account for
$T_c$, our result is that
the vertex correction is not negligibly small at this
temperature,~\cite{AS}\
at variance with previous results.~\cite{AIM,CMM}\
The discrepancy may be due to
a high-energy contribution included in our calculation,
or it is traced back to the above finding of a slight 
renormalization effect on $V_i^{(2)}$.
On the other hand, for $T_c=180$K,
we find that $g^2 =0.73$ in Fig.~\ref{fig:4}
is modified to $g^2 =1.57$.
The large modification in this case
is due to a large coupling constant
to realize that high transition temperature.
The results in this regime must be taken with care.

To the extent that the vertex corrections
that we found for the pairing potentials
are not negligible, the vertex corrections to
eq.~(\ref{Sigma=g2chiG}) should have to be investigated next.~\cite{CMM}\
The latter effect on $\Sigma(\mibp,\mi\omega_n)$
will reduce $T_c$ somewhat
particularly through the pair propagator $|G(\mibpp,\mi\omega_{n'})|^2$
in eq.~(\ref{K(1)}),
according to the above note,
as a result of enhanced quasiparticle damping.
Therefore, 
we will be  ultimately led to
a convergent result of $T_c(g^2)$,
somewhere in between the dashed and solid lines of Fig.~\ref{fig:4}.
The results, however, would then indicate that $T_c\simeq 100$K
is on the verge of practical applicability of this kind of
perturbation theory in $g^2$, as inferred from Fig.~\ref{fig:4}.
Note that, for us in this context,
to suffer a small correction is
more important than to find out a high $T_c$.

In summary, a result of this paper is presented in Fig.~\ref{fig:4},
though the result at high temperature is somewhat
modified as stated above.
Applying perturbation theory to
the eigenvalue of the kernel $K(\mi\omega_n,\mi\omega_{n'})$,
we estimated the vertex corrections to $T_c$
as a function of the coupling constant $g^2$
on the basis of the spin-fluctuation model of
the high-$T_c$ superconductivity.
We found that the effect of Fig.~\ref{fig:3}(b)
is numerically negligible as far as
the $d_{x^2-y^2}$ pairing instability is concerned,
while Fig.~\ref{fig:3}(a) enhances $T_c$ appreciably.
For $T_c\sim 100$K,
the effect of $\Sigma(\mibp,\mi\omega_n)$
mainly comes in through the pair propagator
$|G(\mibpp,\mi\omega_{n'})|^2$,
dressing the vertex functions is not so important.
In a strong-coupling regime at high temperatures,
the vertex corrections become even qualitatively important,
particularly
in case where $T_c$ in the one-loop \'Eliashberg calculation
is substantially suppressed by lifetime effects.


We would like to thank J. Igarashi, M. Takahashi, T. Nagao,
T. Yamamoto and N. Ishimura for valuable discussion.
This work was supported  by
the Japan Society for the
Promotion of Science for Young Scientists.

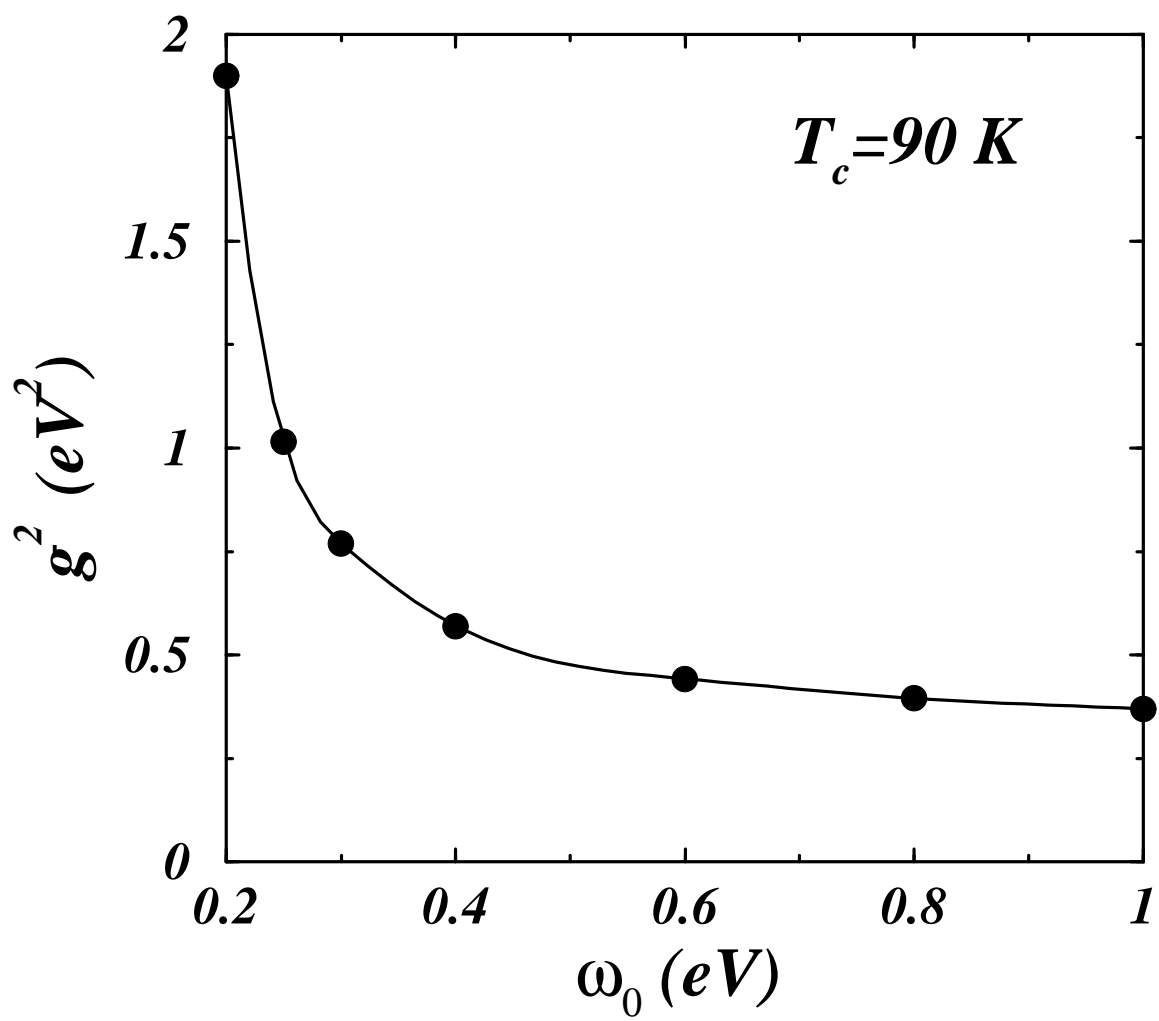
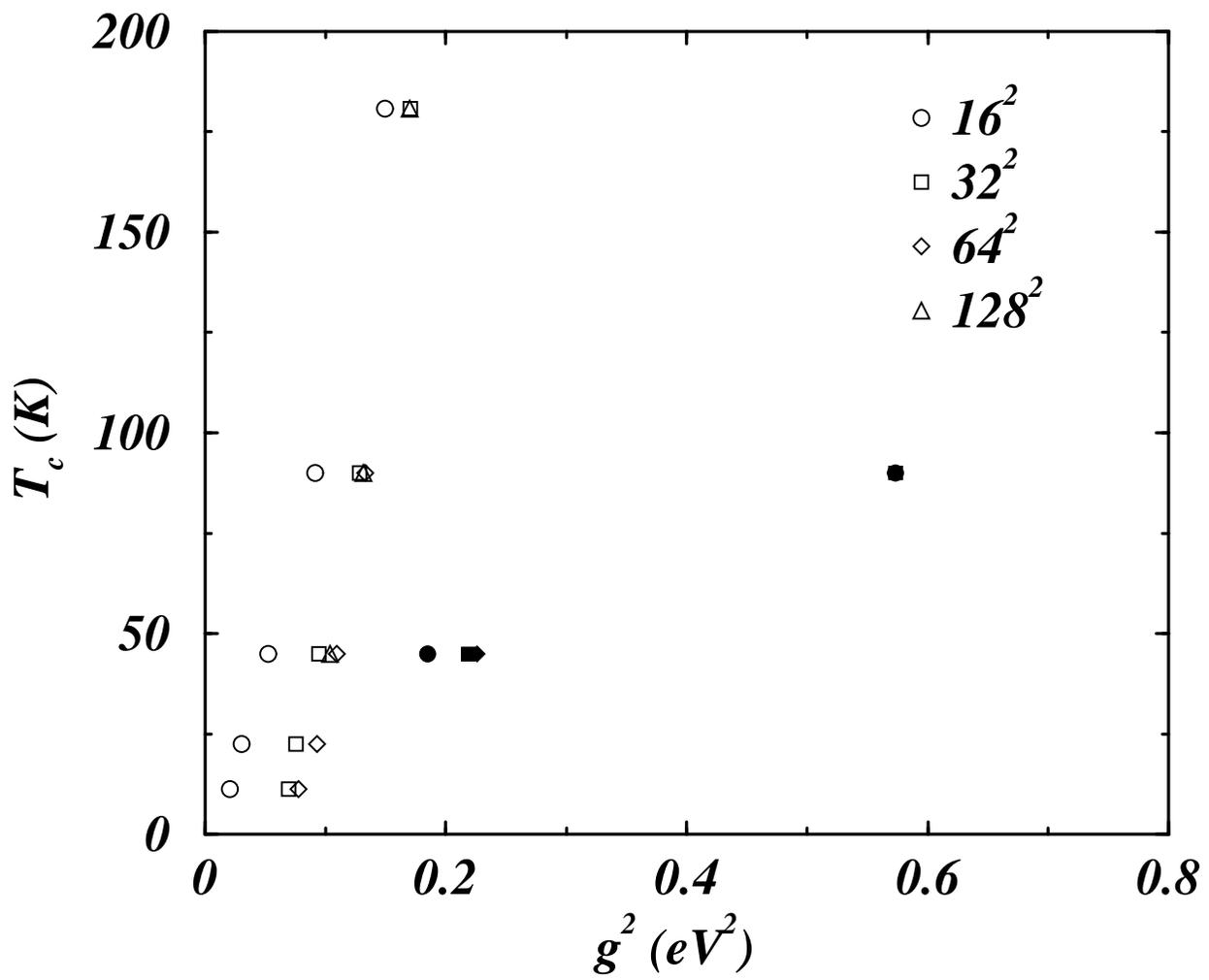
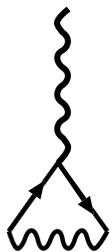
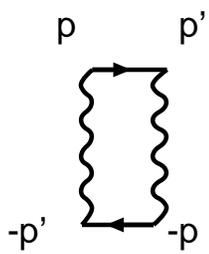
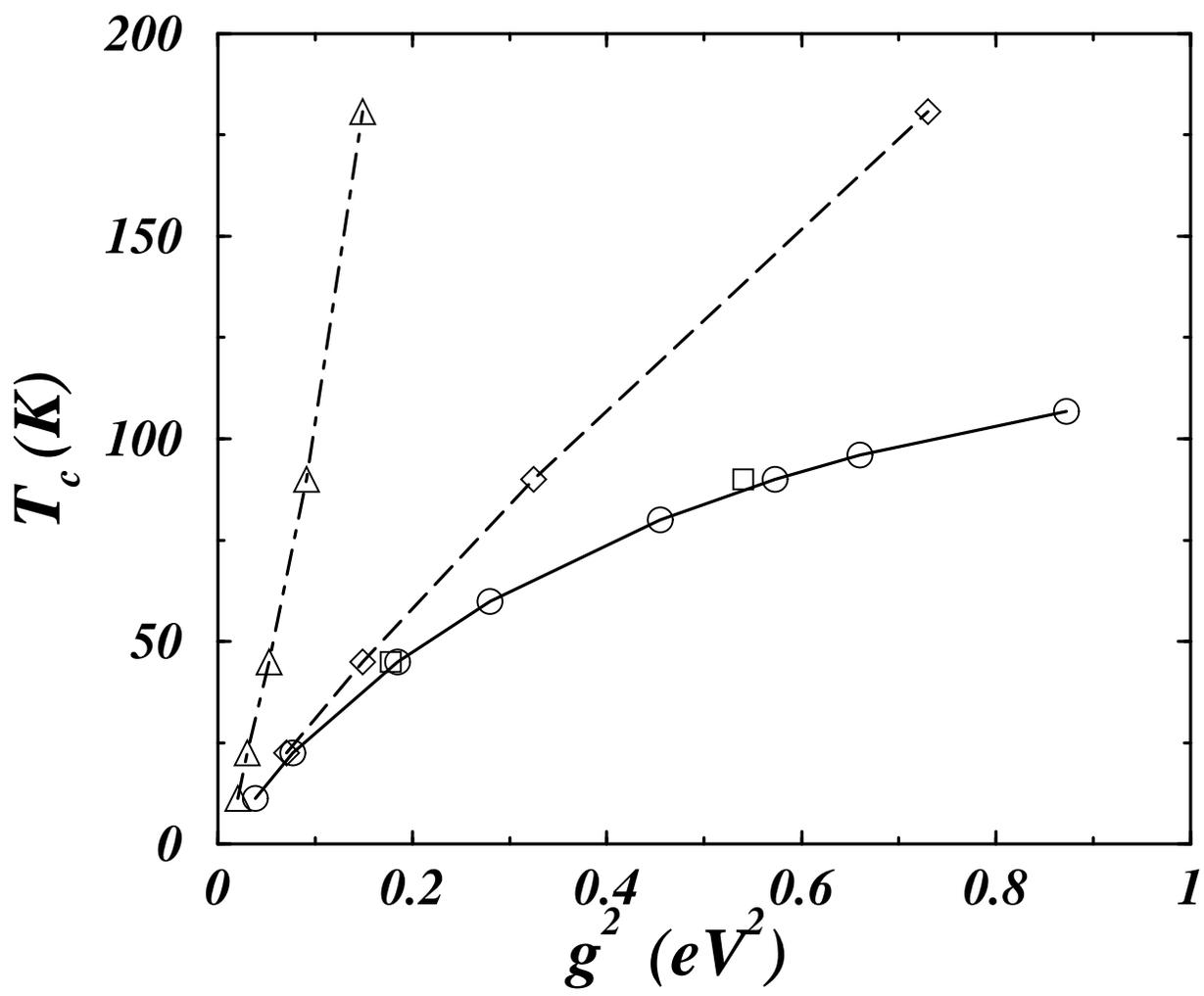
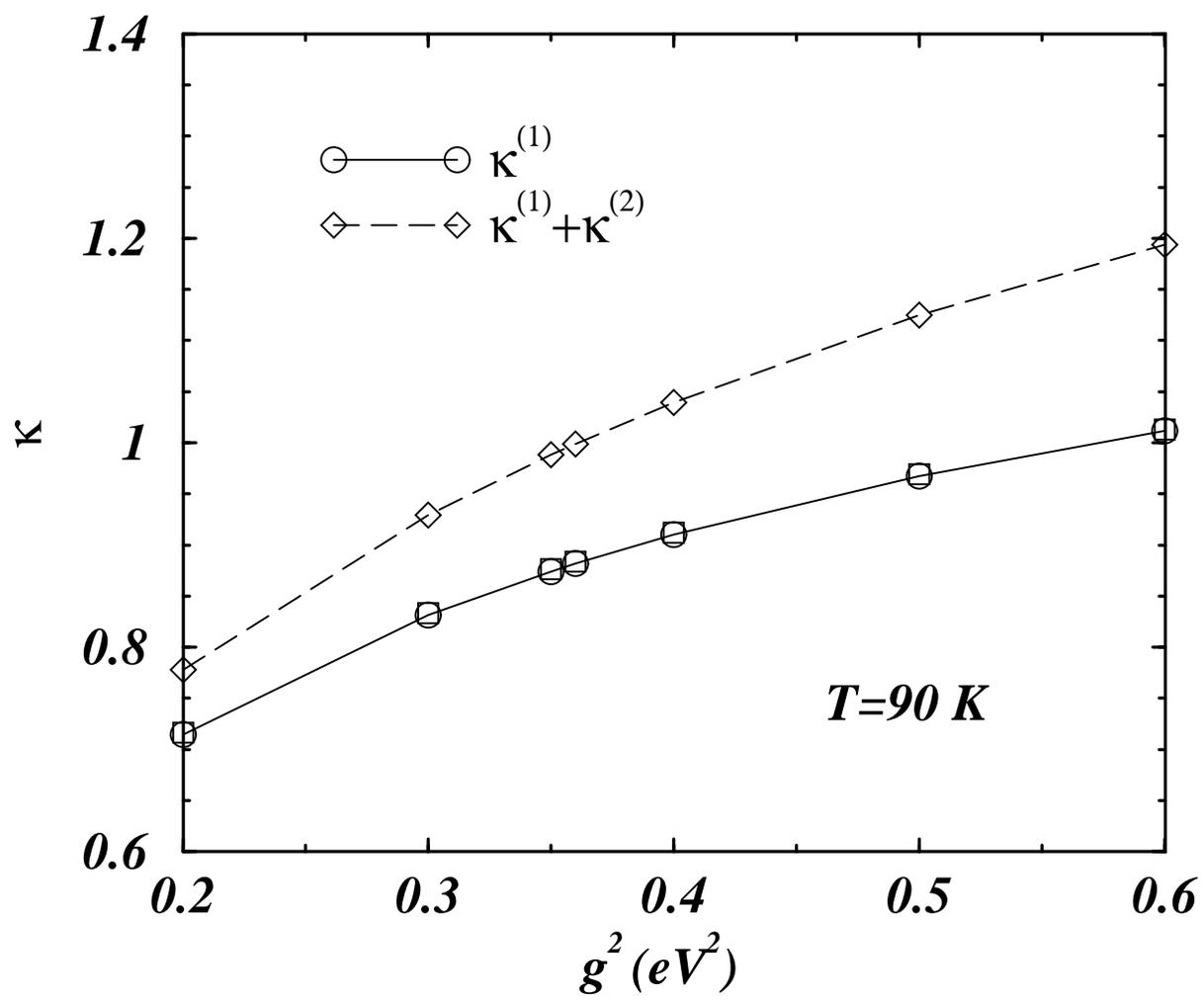
\end{document}